\documentclass[12pt]{iopart}

\newcommand{\lam}{\lambda}

\newcommand{\beq}{\begin{equation}}
\newcommand{\eeq}{\end{equation}}
\newcommand{\ba}{\begin{array}}
\newcommand{\ea}{\end{array}}
\newcommand{\beqa}{\begin{eqnarray}}
\newcommand{\eeqa}{\end{eqnarray}}
\newcommand{\bd}[1]{ \mbox{\boldmath $#1$}  }
\begin{document}
\input{epsf.tex}
\jl{1}

\title[Decay with dissipation]
{Quantum tunneling with dissipation in smoothly joined parabolic potentials}

\author{\c Serban Mi\c sicu\dag
\footnote[3]{e-mail to misicu@theor1.theory.nipne.ro.}}

\address{\dag\ National Institute for Physics and Nuclear
Engineering, Bucharest-M\u agurele, P.O.B. MG-6, ROMANIA}

\begin{abstract}
This paper is devoted to the study of quantum dissipation in cluster
decay phenomena in the frame of the Lindblad approach to quantum open
systems. The tunneling of a metastable state across a piecewise quadratic 
potential is envisaged for two cases : one and two harmonic wells smoothly
joined to an inverted parabola which simulates the barrier.
The width and depth of the second harmonic oscillator well was varied over 
a wide range of values in order to encompass particular cases of tunneling 
such as the double well potential and the cluster decay. 
The evolution of the averages and covariances of the quantum sub-system is 
studied in both under- and overdamped regimes. For a gaussian intial 
wave-packet we compute the tunneling probability for different values of the 
friction coefficient and fixed values of the diffusion coefficients. 
The ansatz used for these coefficients corresponds to the case with temperature 
$T=0$ as happens in the cold fission and cluster decay phenomena. 

\end{abstract}

\submitted


\section{Introduction}

The question of how collective energy is dissipated in various decaying phenomena
occuring in solid-state and nuclear physics is a topic of major interest. 
A particular case is represented by the decay of a metastable state through 
quantum tunneling at very small or zero temperatures. 
Such cold decay phenomena are well known at the microscopic level for nuclei 
undergoing cold fission or emission of heavy clusters \cite{san92,hofman97}
but also for macroscopic systems like Josephson junctions \cite{abrik84} or 
SQUID's \cite{scalap69}. 

Twenty years ago Caldeira and Legget proposed a model of tunneling of a
metastable quantum state at $T=0$ based on the Feynman path integral technique 
\cite{calleg80}. Assuming a linear dependence on the tunneling coordinate $q$ 
for the friction term entering the Lagrangian, they showed that the dissipation 
decreases the tunneling probability.   
Since that time a new formalism for studying quatum dissipative systems emerged.
In this approach the system is {\it opened}, i.e. it is regarded as a subsystem 
interacting with a larger system, referred to as the reservoir or the bath 
\cite{davies76}. The dissipation arises as a consequence of this interaction.

The Lindblad's axiomatic way \cite{lind76} of introducing dissipation in quantum 
mechanics consists in replacing the one-parameter group action 
\beq
U_t(\rho)=
\exp\left (-{\rmi\over\hbar}Ht\right )\rho(t)\exp\left ({\rmi\over\hbar}Ht\right ),
\eeq
of a closed physical system with Hamiltonian $H$, by a new one, 
$\Phi_t(\rho)$ which describes the evolution of the corresponding open 
subsystem. Like $U_t$, this operator should satisfy the conditions of 
self-adjointness ($\Phi_t^*(\rho)=\Phi_t(\rho)$), positivity ($\Phi_t(\rho) > 0$) 
and unitary trace ($\Tr(\Phi_t(\rho))=1$). Moreover
$\Phi_t(\rho)\rightarrow\rho$ when $t\rightarrow 0$.  

Using the above mentioned approach the quantum mechanics of one and 
two coupled harmonic oscillators has been extensively studied 
\cite{sand87a,sand87b,isar94}. Based on these results, very recently, the 
tunneling probability with dissipation was computed exactly for the inverted 
parabolic potential and approximately for the cubic one \cite{adam98,isar99}. 
It was found that there might be cases when friction favours the penetration. 
However the inverted parabola is not a good candidate to describe a real 
tunneling process because no metastable state could be conceived in it. 
The cubic potential is more likely to describe the decay of a metastable state
but in this case the authors faced the problem of dealing with a system of
equations in averages and higher momenta which is not closed. Therefore all
cubic and higher momenta have been droped off providing thus an approximate 
solution.   

In this paper we consider that the quantum decay process, characterized by the
coordinate $q$ is modelled by the tunneling of a one-dimensional potential barrier 
whose explicite form is given by joining first two parabolic segments as in the 
Kramers problem \cite{kram40}, and next we add a third parabolic well beyond the 
barrier.
The description of tunneling phenomena accompanied by dissipation in 
Josephson junctions \cite{melnik91} or fission of heavy compound nuclei
\cite{nix84} offenly employs such barriers. 
Besides computing the averages and covariances of the decay coordinates
we are mainly interested to investigate the dependence of the tunneling 
probability on the friction.

\section{Lindblad approach to open one-dimensional quantum systems}

According to the original paper of Lindblad \cite{lind76} the time
evolution of an observable $A(q,p)$ belonging to a system undergoing a 
non-equilibrium process is written in the Heisenberg picture as
follows
\beqa
\frac{d A}{d t} & = & {\rmi\over \hbar}[H,A]+
{\rmi\lam\over 2\hbar}\left(q[A,p]+[A,p]q-[A,q]p-p[A,q] \right)
\nonumber\\
&-&{D_{qq}\over \hbar^2}[p,[p,A]]
-{D_{pp}\over \hbar^2}[q, [q,A]]
+{D_{qp}\over \hbar^2}\left ( [q, [p,A]] + [p, [q,A]] \right ) 
\label{evol}
\eeqa
where $\lam$ is the friction constant whereas $D_{qq}, D_{pp}$ and $D_{pq}$ are
the quantum diffusion coefficients.
The Hamiltonian $H$ of the one-dimensional subsystem with mass $m$, corresponding
to the above equation of motion, has the general form in coordinate $q$ and 
momentum $p$
\beq
H=\frac{p^2}{2m}+V(q)
\eeq
Note that eq.(\ref{evol}) was derived under the assumption of a weak 
coupling of the subsystem with its environment \cite{talk86}.

Since the dynamics of the subsystem is described in terms of averages and
covariances, for two self-adjoint operators $A$ and $B$ the following
notations are introduced :
\beq
\sigma_A(t) = \Tr (\rho(t)A),~~~~~
\sigma_{AB}(t) = \Tr \left ( \rho(t)\frac{AB+BA}{2}\right )
-\sigma_A(t)\sigma_B(t)
\label{defmedcov}
\eeq
If $A=q$ or $p$ the equations of motion derived from eq.(\ref{evol}) looks
like
\beqa
\frac{\rmd\sigma_q(t)}{\rmd t} & = & -\lam\sigma_q(t) + {1\over m}\sigma_p(t)
\nonumber\\
\frac{\rmd\sigma_p(t)}{\rmd t} & = & -\Tr\left(\rho(t)\frac{\rmd V(q)}{\rmd q}\right)
-\lam \sigma_p(t)
\label{aver}
\eeqa
and
\beqa
\frac{\rmd\sigma_{qq}(t)}{\rmd t} & = & -2\lam\sigma_{qq}(t) + 
{2\over m}\sigma_{pq}(t) + 2D_{qq}
\nonumber\\
\frac{\rmd\sigma_{pp}(t)}{\rmd t} & = & -2\lam\sigma_{pp}(t) - 
\Tr\left(\rho(t)\frac{\rmd V(q)}{\rmd q}p+h.c.\right)
\nonumber \\
&&+ 2\Tr\left(\rho(t)\frac{\rmd V(q)}{\rmd q}\right)\sigma_p(t) + 2D_{pp}
\label{covar}\\
\frac{\rmd\sigma_{pq}(t)}{\rmd t} & = & -2\lam\sigma_{pq}(t) + 
{1\over m}\sigma_{pp}(t)+\sigma_q(t)\Tr\left(\rho(t)\frac{\rmd V(q)}{\rmd q}\right)
\nonumber\\
&&-\Tr\left(\rho(t)\frac{\rmd V(q)}{\rmd q}q\right) + 2D_{pq}
\nonumber
\eeqa
The above set of first-order o.d.e. has been solved exactly for the
harmonic oscillator, $V=m\omega^2q^2/2$, and the inverted parabolic 
potential ($\omega\rightarrow i\omega)$\cite{sand87a,isar99}.

\section{Quantum tunneling across quadratic potentials}

We formulate the decay problem as the tunneling of an initial gaussian
wave packet, confined in a harmonic oscillator well (potential pocket),
located at the left of the barrier, with given values of the position average
$\sigma_{q}(0)$ and covariance $\sigma_{qq}(0)$ :
\beq
\psi(q) = \frac{1}{(2\pi\sigma_{qq}(0))^{1/4}}
\exp\left [-\frac{1}{4\sigma_{qq}(0)}(q-\sigma_{q}(0))^2 +
{\rmi\over\hbar}\sigma_p(0)q\right ] 
\eeq

According to \cite{risken89}, the transition probability for an
irreversible process described by the set (\ref{aver},\ref{covar}) reads
\beqa
\fl W(q,p,t)  = (2\pi)^{-1}({\rm Det}~\bd{\sigma})^{-1/2}
\exp\left \{-{1\over 2}
\left [ \bd{\sigma}^{-1}(t)\right ]_{qq}(q-\sigma_q(t))^2\right. 
\nonumber\\
\lo-  \left. \left [ \bd{\sigma}^{-1}(t)\right]_{pq}
(q-\sigma_q(t))(p-\sigma_p(t)) 
-{1\over 2} \left [ \bd{\sigma}^{-1}(t)\right]_{pp}(p-\sigma_p(t))^2 
\right\}
\eeqa
where $\bd{\sigma}$ is the 2$\times$2 matrix of the covariances 
defined in eq.(\ref{defmedcov}).
The probability to find the wave-packet to the right of the barrier is 
\cite{weiner69}:
\beq
P(q_b;t) = \int_{q_b}^{+\infty}dq\int_{-\infty}^{+\infty}dp~W(q,p,t)=
{1\over 2}{\rm erfc}\left ( \frac{\sigma_q(t)-q_b}
{\sqrt{2\sigma_{qq}(t)}}\right )
\label{probtun}
\eeq
The decay rate, or fission width, is given by the ratio between the diffusion 
current across the barrier, $J(q_b;t)=\int_{-\infty}^{+\infty}dp~W(q_b,p;t)$ and 
the tunneling probability
\beqa
 \Gamma_f(t)   =  \frac{J(q_b;t)}{P(q_b;t)}\nonumber\\
\lo= {1\over \sqrt{2\pi\sigma_{qq}(t)^3}}
\left (\sigma_{qq}(t)\sigma_{p}(t)+\sigma_{pq}(t)(q_b-\sigma_q(t))\right)
\frac{\exp\left [{- \frac{(\sigma_q(t)-q_b)^2}{2\sigma_{qq}(t)}}\right ]}
{{\rm erfc}\left ( \frac{\sigma_q(t)-q_b}
{\sqrt{2\sigma_{qq}(t)}}\right )} 
\eeqa

The purpose of the next two subsections is to develope the above formalism
for two case studies, namely the tunneling of a metastable state from
a parabolic pocket to a quadratic unbound potential and next the tunneling 
between two quadratic wells with finite depths, separated by an inverted parabola.  

\subsection{Two smoothly joined parabolic potentials}

In this case the potential is given by joining at $q_t$, an inverted oscillator 
to an upright ground-state oscillator (see Fig.1):
\beqa
V(q) & = & V(q_a) + {1\over 2}\Omega_1^2(q-q_a)^2,~~~q\leq q_t
\nonumber\\
& = & V(q_b)-{1\over 2}\Omega_b^2(q_b-q)^2,~~q\geq q_t
\eeqa
In the above expression one suppose that the height of the barrier 
$B=V(q_b)-V(q_a)$, and the frequency $\Omega_b$ of the inverted oscillator 
are known. 
In the case of nuclear reactions this means to know the
barrier from the heavy-ion interaction of the two clusters or the 
dependence of the total energy on deformation \cite{sat79,quent78}.
Assigning a decay energy $E_0$ to the particle of mass $m$ we suppose
that the minimum of the first harmonic oscillator well $q_a$ coincides with the 
first turning point.  
Naturally, since we deal with smoothly joined segments, the continuity of
$V$ and its derivative $\rmd V/\rmd q$ is required at $q=q_t$.
From here we get the frequency of the well's bottom :
\beq
\Omega_a^2 = \frac{2\Omega_b^2B}{\Omega_b^2(q_b-q_a)^2-2B}
\eeq
and the position of the joining point :
\beq
q_t = \frac{q_a\Omega_a^2+q_b\Omega_b^2}{\Omega_a^2+\Omega_b^2}
\eeq

In order to integrate the systems (\ref{aver},\ref{covar}) we need
to establish the initial conditions. The initial wave-packet will
be confined inside the harmonic oscillator well centered at $\sigma_q(0)$.
In our calculation, motivated by specific examples from nuclear fission
we choosed $\sigma_q(0) = q_a$, i.e. the gaussian wave-packet is centered at the
minimum of the potential pocket. In order to overcome the barrier the initial 
momentum should satisfy the inequality $ \sigma_p(0)> \sqrt{2mB}$ 
for a vanishing friction coefficient $\lam$. The initial value of the coordinate 
covariance $\sigma_{qq}(0)$ is choosen in such a way that its time derivative 
vanishes. 
Assuming that $\sigma_{pq}(0)$ vanishes too, we then get from eq.(\ref{covar})
\beq
\sigma_{qq}(0)=\frac{D_{qq}}{\lam}
\eeq
Since we deal with a gaussian wave packet, the momentum and the coordinate
covariances are simply related by means of the following relation \cite{isar99}
\beq
\sigma_{pp}(0) = \frac{\hbar^2}{4\sigma_{qq}(0)}
\eeq

For the diffusion coefficients we take the rotating-wave approximation
\cite{jang86} 
\beq
D_{qq}=\frac{\lam\hbar}{2\sqrt{m\Omega_a^2}}D_{pp}=m\Omega_a^2~D_{qq},~~~~~D_{pq}=0
\eeq
These values were obtained from eq.(\ref{covar}) by asuming that the initial 
Gaussian metastable state is a Gibbs state at $T=0$.

In Figs.2 and 3 we represented the averages and covariances of the coordinate and 
momentum taking four choices of the friction coefficient $\lam$. We assumed that 
the initial momentum is the same, i.e. $\sigma_p(0)$=1200 MeV. The general trend
is a damping of the motion due to friction. Above a critical value $\lam_{cr}$ of 
the friction coefficient all the momenta tends to infinity, faster if $\lam=0$ 
and slower if $\lam\nearrow\lam_{cr}$. Bellow this critical value the gaussian 
wave packet will perform damped oscillations around the potential minimum.
From the inspection of Figure 2 it is obvious that the wave-packet is spending
a longer time in the barrier if $\lam$ is approaching its critical value.   
The decreasing of the tunneling probability (see eq.(\ref{probtun})) with $\lam$ 
is shewn in Fig.4. After a certain transition time ($\approx 5\times 10^{-22}$s)
the tunneling probability will reach a non-vanishing asymptotic value for 
$\lam<\lam_{cr}$. Looking on the first panel of Fig.2 one sees that after this 
lapse of time the centroid of the gaussian is located on the other side of the 
barrier. Consequently, in the present approach, the tunneling time, a different
quantity from the metastable state life-time, is assigned to the time necessary
to attain an asymptotic value $P(q_b;\infty)$. For a discussion 
of this subject in the frame of the WKB-approximation and time dependent 
approach to Schr\"odinger equation, see \cite{landau94,miscar99}.     

For $\lam>\lam_{cr}$ the tunneling probability eventually tends to zero but
at the beginning of the motion, between certain lapses of time, the right tail of 
the wave function is found inside the barrier. If the initial momentum 
$\sigma_p(0)$ would be increased then the barrier is once again overcomed.
However in tunneling phenomena like cluster decay the initial average momentun of
the wavefunction should not exceed certain limits and thus upon comparison 
with the penetrabilities deduced from experiment, the range of possible values 
of the friction coefficient could be deduced.

Thus, for two smoothly joined parabolas, the friction reduces the tunnelling
probability of the metastable state in the frame of Lindblad's axiomatic
approach.

\subsection{Three smoothly joined parabolic potential }

The case studied bellow suffers from the deficiency that the right tail of
the potential falls to $-\infty$. In usual applications we deal with tails
which tends to $+\infty$ or to 0.
We thus join a third parabolic segment at a second turning point,
$q_{t2}$, centered at $q_c$ (see fig.5)  :   
\beqa
V(q) & = & V(q_a) + {1\over 2}\Omega_a^2(q-q_a)^2,~~~q\leq q_{t1}
\nonumber\\
& = & V(q_b)-{1\over 2}\Omega_b^2(q_b-q)^2,~~q_{t1}\leq q\leq q_{t2}\\
& = & V(q_c)+{1\over 2}\Omega_c^2(q-q_c)^2,~~ q\geq q_{t2}
\nonumber
\eeqa 
Imposing the same conditon of continuity of $V$ and its derivative $\rmd V/\rmd q$
at  $q=q_{t2}$ one gets
\beq
\Omega_c^2 = \frac{2\Omega_b^2 \Delta V_{bc}}{\Omega_b^2(q_c-q_b)^2-2\Delta V_{bc}}
\eeq  
and
\beq
q_{t2}=\frac{q_c\Omega_c^2+q_b\Omega_b^2}{\Omega_c^2+\Omega_b^2}
\eeq
where $\Delta V_{bc}=V(q_b)-V(q_c)$ is the difference between the top of the
barrier and the bottom of the second well.

In Fig.6 the evolution of averages were represented 
for two harmonic wells with equal bottoms, i.e. $V(q_b)=V(q_c)$ (a)
and $V(q_c)=0$ (b). For the location of the second well minima several values 
were selected. The first one ($q_c=16.5 $ fm) corresponds almost to the  
symmetric double-well problem.  In the absebce of friction the tendency of 
$\sigma_q(t)$ is to oscillate indefinetely between these two wells with amplitude 
and period determined by the location of the second minimum $q_c$. Like in the 
case examined in the preceeding subsection, a critical value of $\lam$ will occur. 
Above this value the wave-packet will never cross the barrier and will perform 
damped vibrations inside the first well.  If $\lam \leq \lam_{cr}$ then the 
gaussian will pass in the second well and will not return in the first well due to 
the kinetic energy loose by friction. In this last case the centroid of the wave 
packet will tend asymptotically to the center of the second well $q_c$. Ase one 
sees from Fig.6 there are no qualitative difference between the motions in the 
potential with  $V(q_b)=V(q_c)$ and $V(q_c)=0$. 

In terms of tunneling probability, the results obtained for the three smoothly
joined parabolas are displayed in Fig.7. If friction is absent the tunneling 
probability will tend to 0.5, i.e. half of the wave-packet is found at the 
right of the barrier's top, $q_b$, and half to its left. If friction is switched
on, but up to $\lam_{cr}$, then asymptotically the wave-packet tunnels 100\%
in the second well after a certain lapse of time. This lapse of time will be
shortened if the minima of the second well is shifted towards higher values. 
When $\lam_{cr}$ is exceeded the tunneling probability  goes to zero since the 
wave packet is constrained by the high friction to not leave the first harmonic 
well. 

Contrary to the preceeding case, for the double-well problem the value of the 
friction constant, bellow a threshold, favours the tunneling. Above this
threshold the tunneling is totally hindered.

\section{Conclusions}

In this paper the study of tunneling, in a one-dimensional piecewise potential 
built-up from two and three parabolic segments, 
was carried out in the frame of the Lindblad approach to quantum systems 
for a gaussian wave-packet.
In this way the interaction of the system with the environment was taken into
account and the effect of dissipation incorporated in the equation of motion
for the averages and the covariances. 
In view of the applications in cold fission
and cluster decay only the case with temperature $T$=0 was considered.
This amounts to admit a linear dependence of the diffusion coefficients on the 
friction constant for an initial metastable state. 

For the two-smoothly joined parabolic potentials, if the friction constant 
is bellow a certain critical value, the gaussian wave-packet cross
just one time the barrier and afterwards it moves to infinity with a high or 
a low momentum, depending on how large is the friction. 
The computation  of the tunneling probability in this case confirms older 
results \cite{calleg80} which claimed that the effect of linear friction decreases 
the chance to find the wave-packet beyond the barrier.

In the case of three-smoothly joined parabolas the situation is different.
If the motion is unviscous then the wave packet will oscillate unhindered 
between the infinite walls of the two wells. Therefore, the probability current 
flowing from the left to the right of the barrier will be compensated by the one 
flowing in the opposite direction. If the friction is present in the 
system, although bellow a certain limit, the tunneling is stimulated, in the 
sense that asymptotically  all the wave-function is located in the right well.
This happens no matter how large or deep is this second well. Only the time 
necessary to establish the equilibrium will depend on the features of the employed 
potential.

It would be interesting to extend the calculations presented in this paper 
to other one-dimensional potentials which are more likely to simulate the 
behaviour of systems undergoing decay by quantum tunneling. 
One of the limitations of the present approach is that the potential needs to be 
provided in analytical form, preferably as a polynomial of lower degree. 
Already for a cubic polynomial complications occur because the system of first
order differential equations describing the time evolution of averages and higher 
momenta is non-linear and is not closed. A solution would be to consider the 
realistic one-dimensional potential as a smooth piecewise curve given by cubic 
splines. Otherwise, the exact treatment of the problem would be to solve directly
the time-dependent Schr\" odinger equation, which unfortunately is non-local
and non-linear\cite{sand87a}.  

\ack I am very gratefull to dr.A.Isar and Prof.A.S\u andulescu for revealing me 
many facets of the quantum theory of open systems.

\newpage
\Figure{Gaussian wave-packet with average initial momentum $\sigma_p(0)$
centered at $q_a$ undergoing tunneling across a barrier of height $B$ 
built-up from two smoothly joined parabolas.}
\Figure{The time dependence of the average values of the coordinate 
$\sigma_q(t)$ and momentum $\sigma_p(t)$ for two smoothly joined parabolas.
Four different values of the friction constant $\lam$ are taken.}
\Figure{The evolution of $\sigma_{qq}(t)$, $\sigma_{pp}(t)$ and $\sigma_{pq}(t)$ 
for friction constants taken as in Fig.2.}
\Figure{The decrease of the tunneling probability in time with the friction
constant in the two smoothly joined parabolas case.}
\Figure{Same as in Fig.1 but for three smoothly joined parabolas. The location 
$q_c$ and depth of the second well minima are varied over a wide range.}
\Figure{The evolution of $\sigma_q(t)$ for three smoothly joined parabolas.
Three different values are taken for the friction constant $\lam$ and four
choices for $q_c$. In (a) the depth of the two harmonic wells are equal, 
i.e. $V(q_a)=V(q_b)$, whereas in (b) $V(q_c)=0$.}
\Figure{The tunneling probability in the three smoothly joined parabolas case
for (a) $V(q_a)=V(q_b)$ and (b) $V(q_c)=0$.}

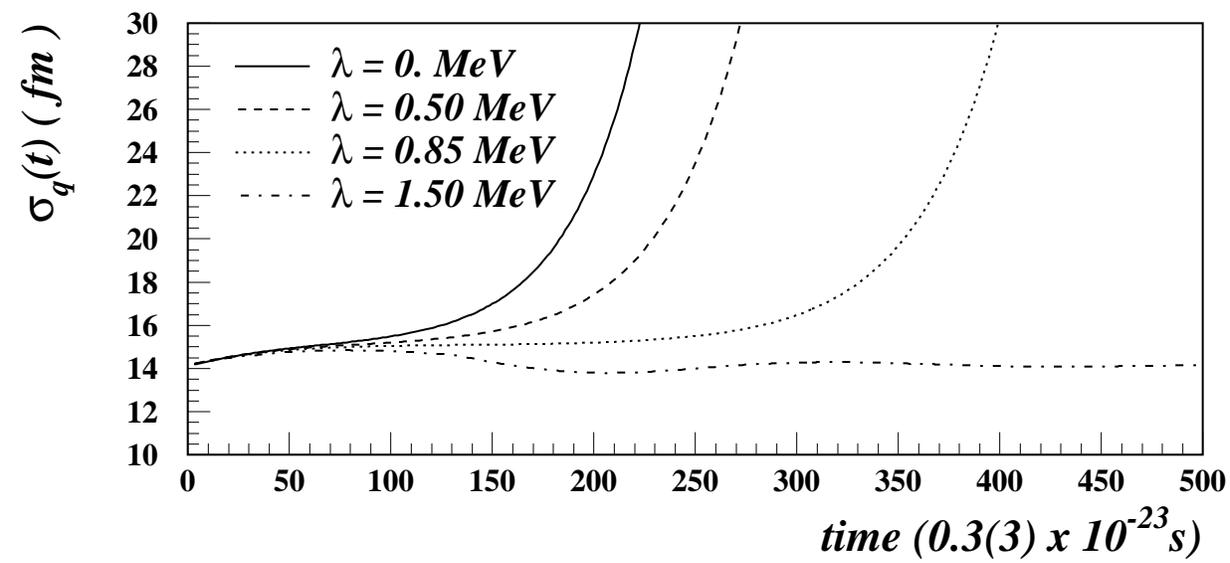
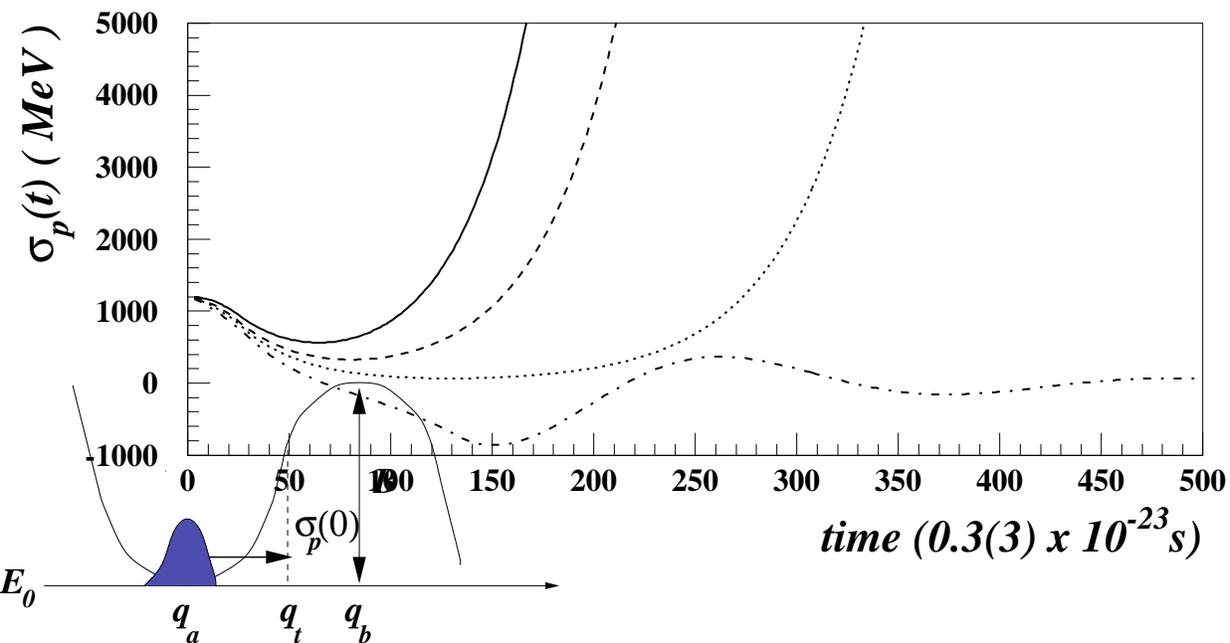
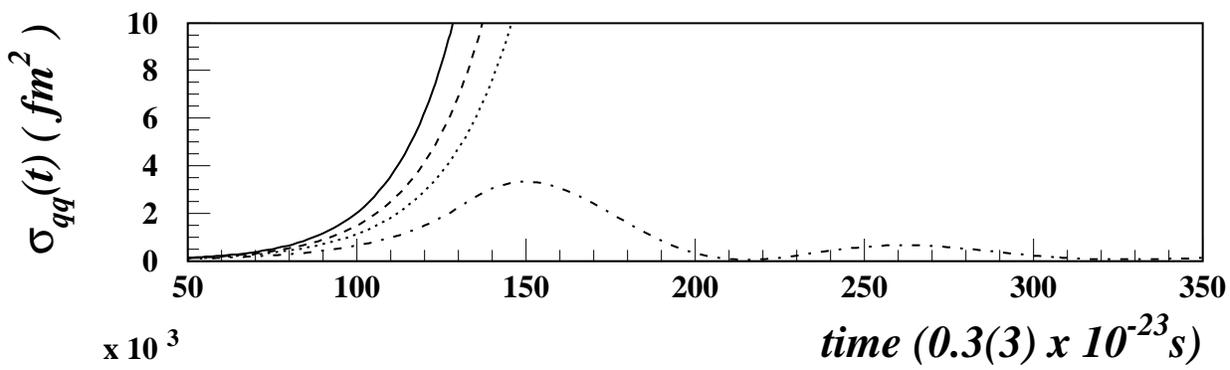
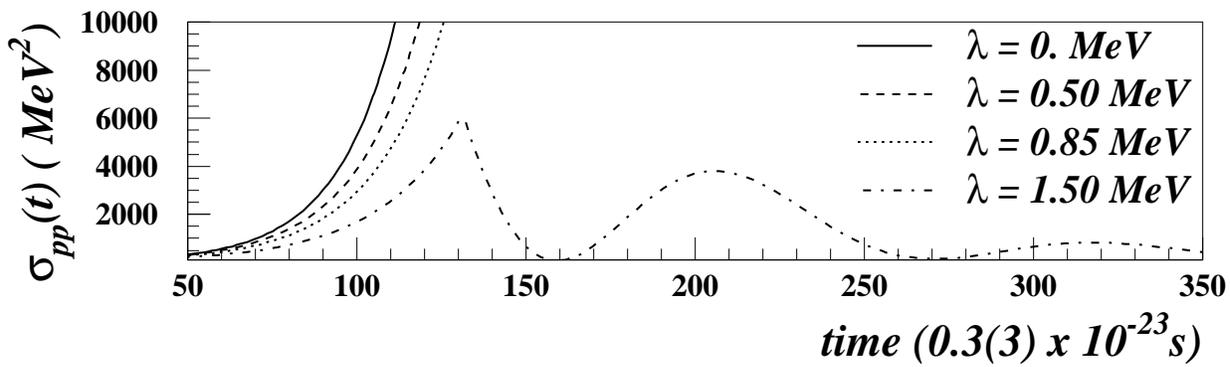
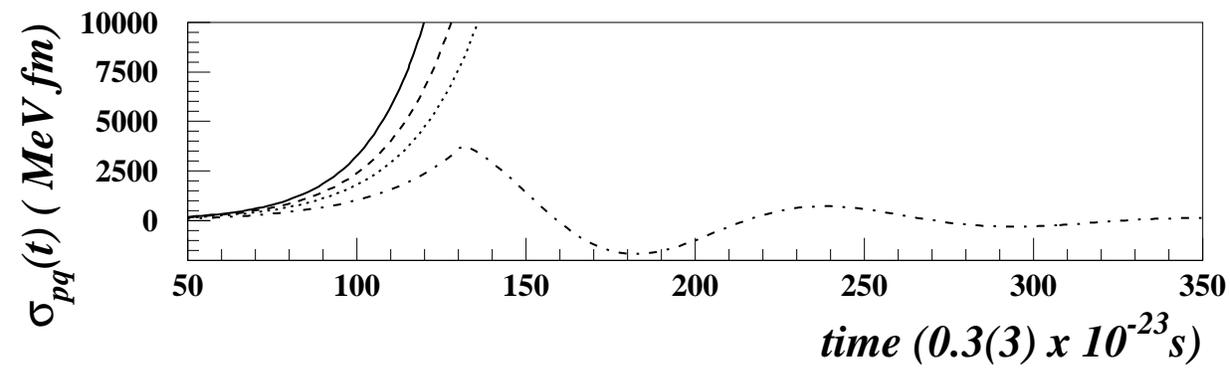
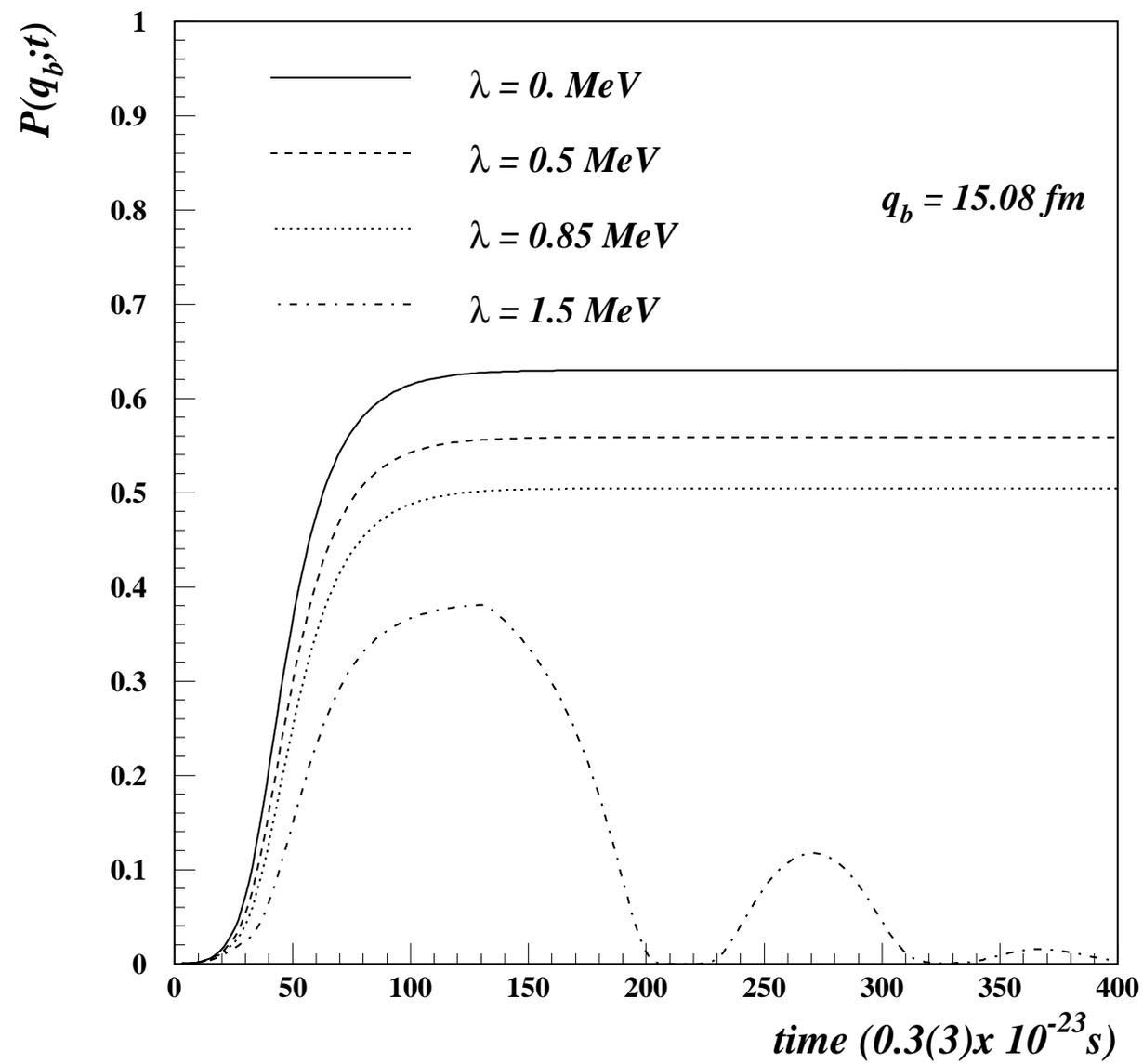
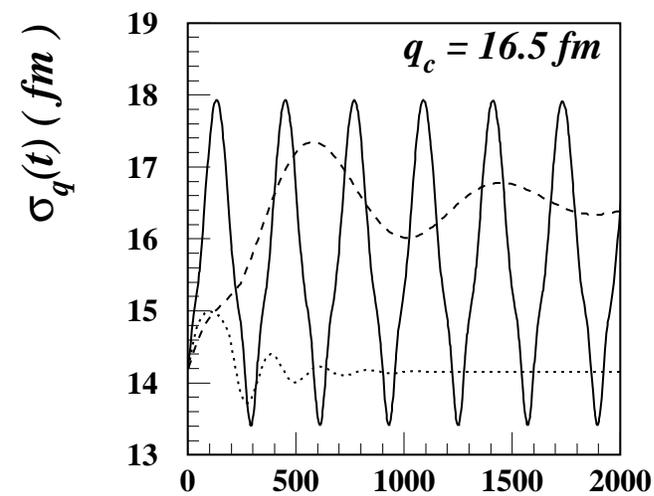
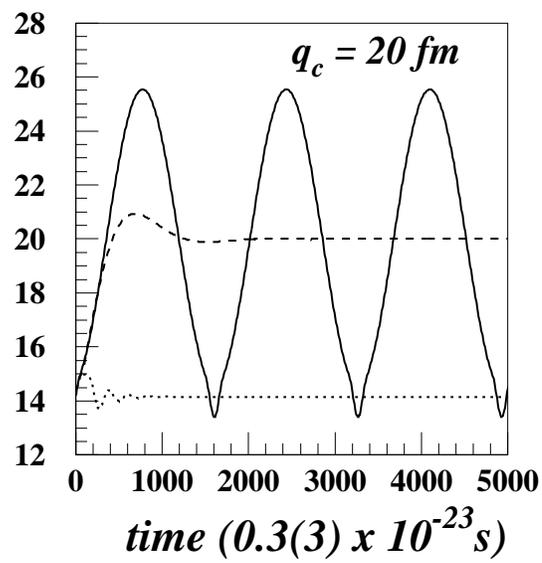
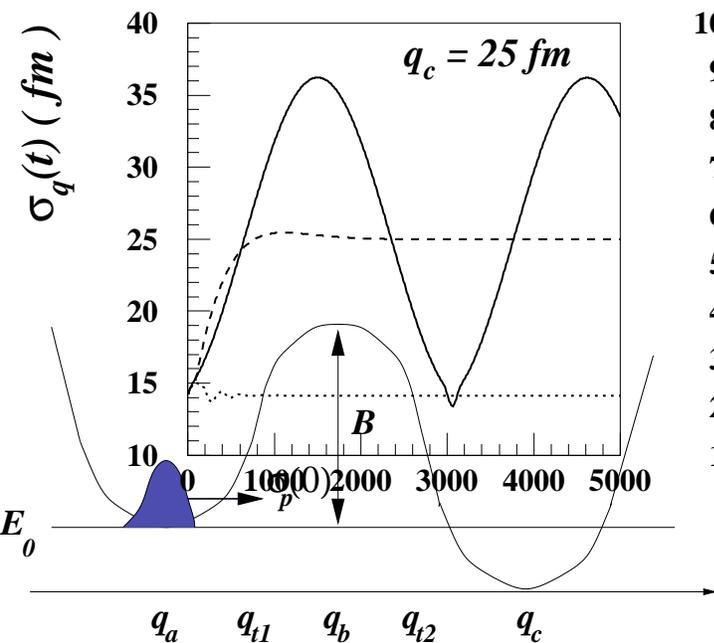
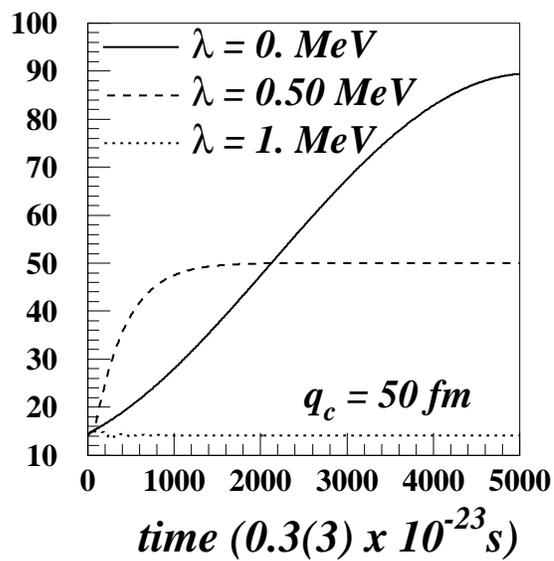
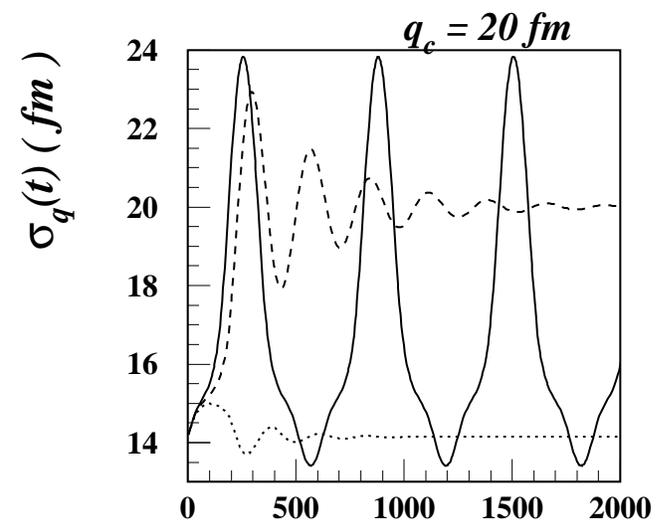
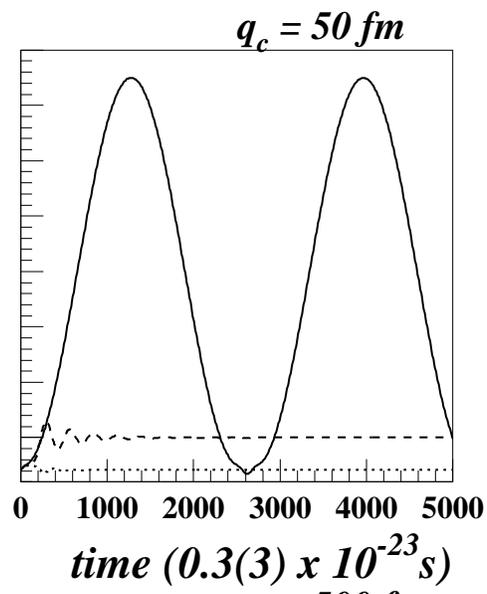
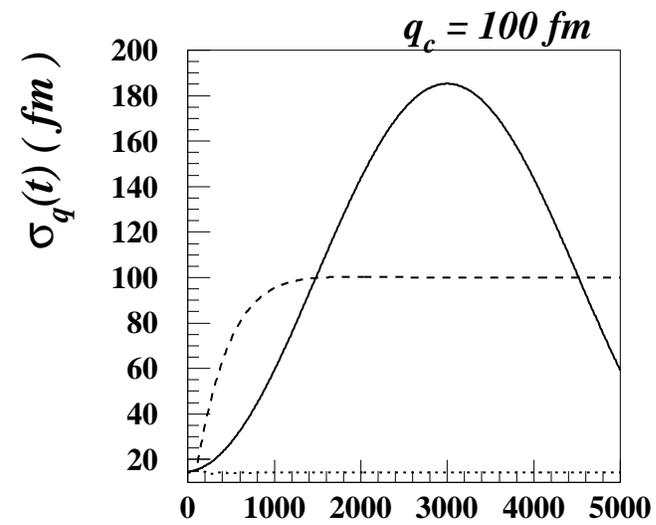
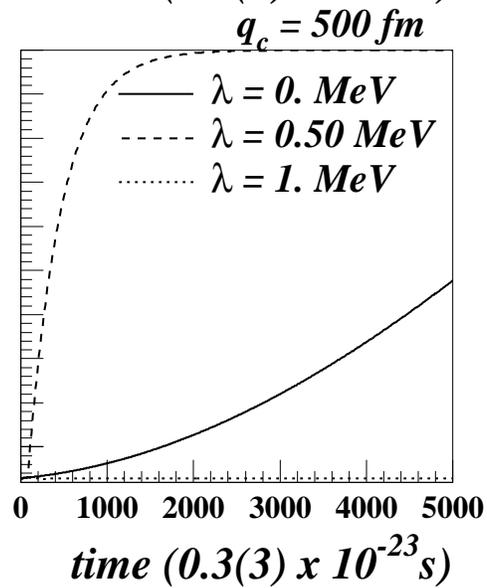
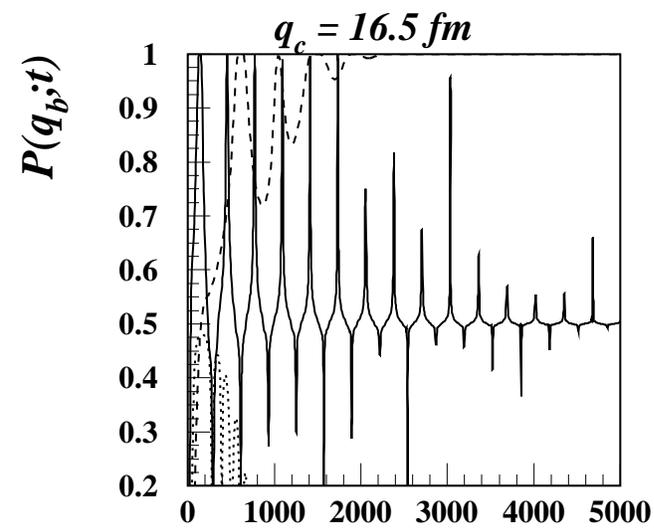
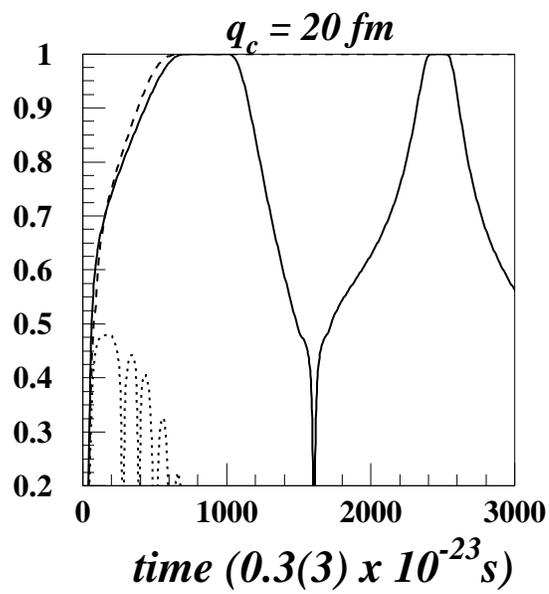
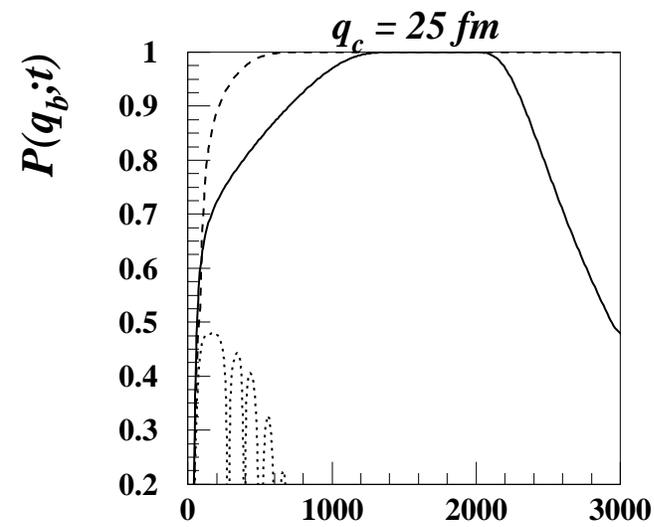
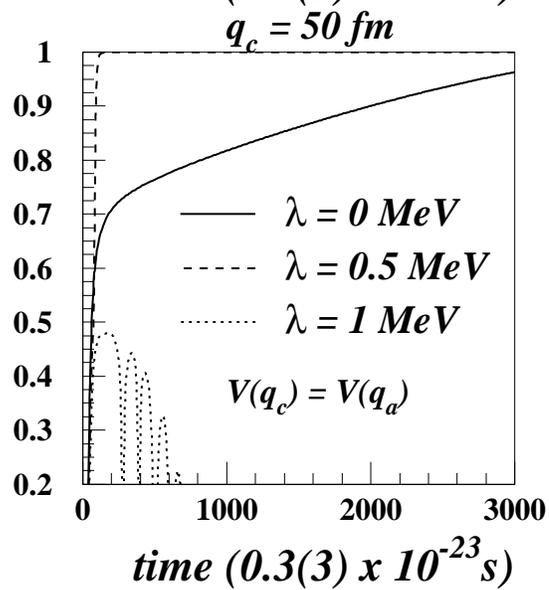
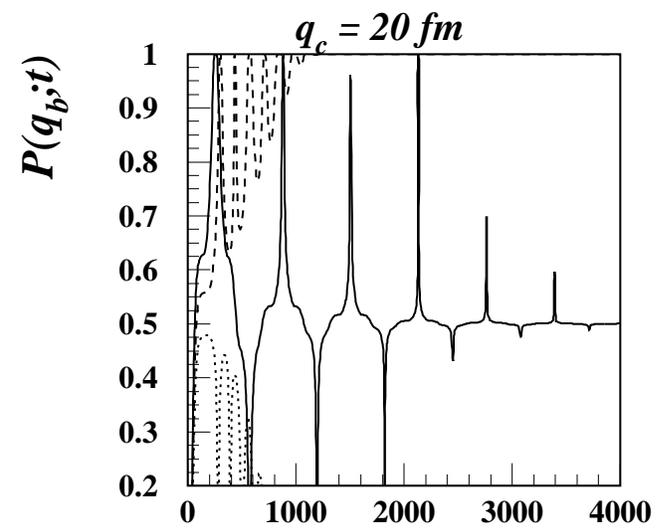
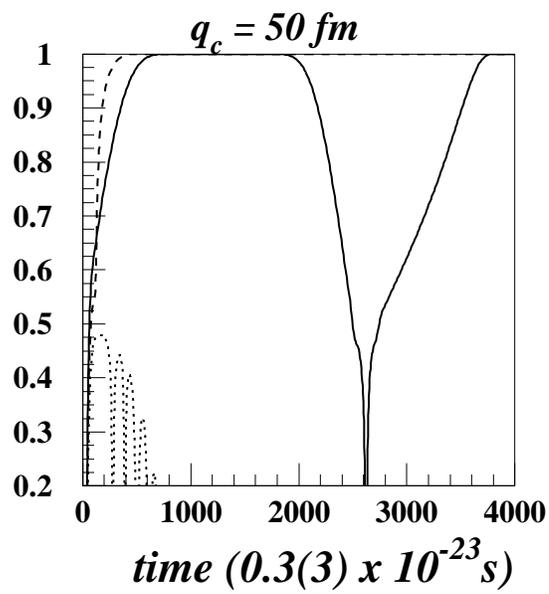
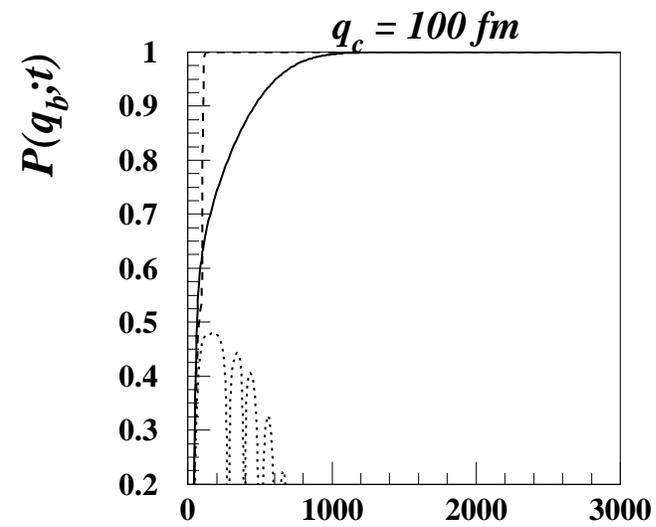
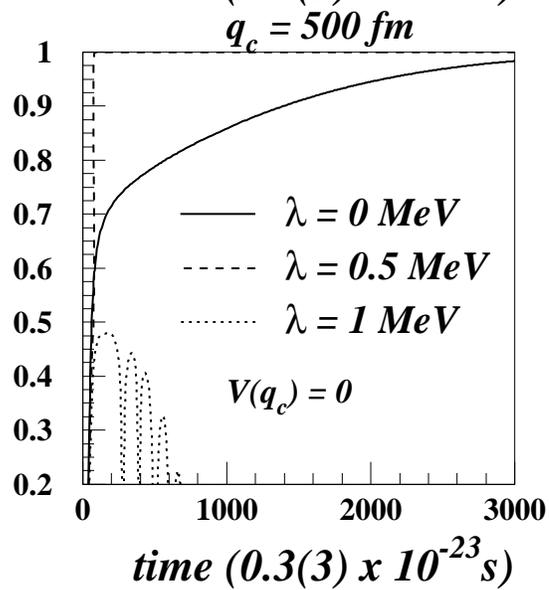
\end{document}